\documentclass[reprint,amsmath,showpacs,amssymb,aps]{revtex4}
\usepackage{graphicx,color}
\newcommand{\Array}[2]{\left(\begin{array}{#1}#2\end{array}\right)}

\begin{document}

\title{CKM and PMNS mixing matrix from $SO(2)$ flavor symmetry}
 
\author{Guojun Xu$^a$, 
Ying Zhang$^{a,b}$\footnote{E-mail: hepzhy@mail.xjtu.edu.cn. Corresponding author. }}
\address{$^a$School of Physics, Xi'an Jiaotong University, Xi'an, 710049, China}

\address{$^b$Institute of Theoretical Physics, Xi'an Jiaotong University, Xi'an, 710049, China}

\begin{abstract}
The relation between quark masses and CKM mixing is studied based on  an approximate chiral $SO(2)_L\times SO(2)_R$ flavor symmetry of quark mass matrix.  In mass hierarchy limit, the mass ratio effect to CKM mixing is suppressed, which separates mass hierarchy and quark flavor mixing into two independent problems. We show that CKM mixing is dominated by two left-handed $SO(2)_L^{u,d}$ symmetry while mass hierarchy only provides   slight corrections.  The same mixing structure is  generalized to lepton sector with extended Dirac neutrinos. The common  flavor mixing 
provides a novel comprehension on the relation between quark CKM mixing and lepton PMNS mixing.

\end{abstract}

\maketitle

	After Higgs particle was found in 2012, all predictions of  the Standard Model (the SM) has been realized \cite{2012Higgs}. However,  some flavor problems on mass patterns and flavor mixings of quarks and leptons still keep their mysteries \cite{2019Zupan,Xing2020PR}. 
In the SM, quark mass matrix $M^q$ stems from   Yukawa couplings after electroweak symmetry breaking.
By performing a bi-unitary transformation $U_L^qM^q(U_R^q)^\dag$, $M^q$ can be diagonalized to hierarchal masses ${\rm diag}[m_1^q,m_2^q,m_3^q]$ with  $m_1^q\ll m_2^q \ll m_3^q$.  At the same time, left-handed unitary transformations $U_L^{u}$ and $U_L^{d}$   involve quark flavor mixing in charged current weak interactions, i.e.  it leads to quark CKM mixing matrix $V_{CKM}=U^u_L(U^d_L)^\dag$.
However, unencrypted Yukawa couplings leave mass hierarchy and flavor mixing  unknown flavor puzzles.
The same case  exists in the lepton sector with massive Dirac neutrinos. 
At present it is not clear  whether these two flavor problems are controlled by independent mechanism, or one dominates another.
Therefore, decoding mass pattern is a key to recover flavor nature. 

In  well-known researchs \cite{FritzschNPB1979,FritzschPLB1986,Fritzsch1987PLB}, CKM mixing  was closely related to  quark mass hierarchies.  The reason  is that  CKM mixing matrix has a hierarchal structure, i.e. three mixing angles are all relatively small. But in the lepton sector, instead of a hierarchal structure, PMNS matrix shows an approximate $\mu-\tau$ symmetry as its characteristics \cite{XingarXiv221011922,ChenPRD2004}, which demands a different mass pattern from quarks. 
However, inspired by the similarity of quark and lepton mixings, some physicists  insist on a common mass pattern hidden 
behind these ambiguous Yukawa couplings.
With this goal, this paper  focuses on a general flavor mixing structure from a common flat mass matrix. Under mass hierarchy, it is found out that a chiral $SO(2)_L\times SO(2)_R$ symmetry dominates CKM/PMNS mixing structure. We then verify this mixing structure by fitting quark and lepton mixing parameters.   As a common flavor  structure for both quarks and leptons, these results are conducive to revealing  the nature of flavor in the future.

An earlier work  \cite{FritzschNPB1979} proposed that CKM mixing angles were dominated by quark mass ratio based on a special mass pattern. 
After phase redefinitions of quark fields, CKM mixing matrix was expressed in  a factorized form
	\begin{eqnarray}
		V_{CKM}=R^u\Array{ccc}{1 && \\ & e^{i\lambda_1} & \\ && e^{i\lambda_2}}(R^d)^T
		\label{eq.Vckmstruc0}
	\end{eqnarray}
with phases $\lambda_1$ and $\lambda_2$. 
$R^{q}$ (for $q=u,d$) is a $SO(3)$ rotation that diagonalizes real mass matrix $M^q$
	\begin{eqnarray}
		R^{q}M^{q}(R^{q})^T={\rm diag}\left(m_1^{q},m_2^{q},m_3^{q}\right)
		\label{eq.RMR}
	\end{eqnarray} 
The similar factorized structure  was  also proposed from polar decomposition of a  complex  matrix in  \cite{ZhangNPB2022}. It hints that Yukawa interaction can be rewritten in a Yukawa basis in which Yukawa couplings become real. And complex phases required by non-vanishing CP violating angles can be explained from a transformation between gauge basis and Yukawa basis.
In \cite{FritzschNPB1979}, to address hierarchal quark masses,  a quark mass pattern was assumed in terms of  sequential see-saw mechanism
	\begin{eqnarray}
	M^{q}=m_3^q\Array{ccc}{0 & \epsilon_1^{q} & 0 \\ \epsilon_1^{q} & 0 & \epsilon_2^{q} \\ 0 & \epsilon_2^{q} & 1}
	\label{eq.seqseesawmatrix}
	\end{eqnarray} 
In this pattern, the third generation transfers its mass to the second by a small  correction $\epsilon_2^{q}$, and the second transfers its mass to the first  by $\epsilon_1^{q}$ in the same way.
This matrix can be diagonalized by a two-step rotation	
	\begin{eqnarray}
		R^q=R_3(\phi_1)R_1(\phi_2)
		\label{eq.OldRq}
	\end{eqnarray}
Here, $R_i$ is a 3-dimensional real rotation along the $i$-th axis.
Rotation angle $\phi_2$ is responsible to diagonalizing $2\times 2$ block matrix with non-diagonal element $\epsilon_2^{q}$ that generates the second generation mass $m_2^q$. And $\phi_1$ is responsible to diagonalizing block matrix with non-diagonal element $\epsilon_1^{q}$ to generate $m_1^q$.
 They are given by simple formulae
 	\begin{eqnarray}
	\tan \phi_2=\sqrt{m^q_2/m^q_3}=\sqrt{h_{23}^q}
	\\
	\tan \phi_1=\sqrt{m^q_1/m^q_2}=\sqrt{h_{12}^q}
	\end{eqnarray}
Here, hierarchy $h_{ij}^q$ is defined by $h_{ij}^q\equiv m_i^q/m_j^q$ with generation index $i,j$.
In this way, quark mass hierarchies are encoded into  CKM matrix.

However, Eq. (\ref{eq.OldRq}) neglects a rotation symmetry in CKM mixing.
In mass basis, quark mass matrix  has  hierarchal eigenvalues
	\begin{eqnarray}
		m_3^q{\rm diag}\Big(h_{12}^{q}h_{23}^q ,  h_{23}^q ,1\Big)
	\end{eqnarray}
Under mass hierarchy limit $h_{12}^q=h_{23}^q=0$, 
above diagonal matrix   becomes $m_3^q{\rm diag}(0,0,1)$, which is obviously invariant under a chiral $[SO(2)_L\times SO(2)_R]^q$ symmetry 
	\begin{eqnarray}
		R_{3}(\theta^q_L){\rm diag}(0,0,1)R_{3}^T(\theta^q_R)={\rm diag}(0,0,1)
	\end{eqnarray}
Since sequential seesaw pattern $M^{q}$ in Eq. (\ref{eq.seqseesawmatrix}) tends to $m_3^q{\rm diag}(0,0,1)$ when $\epsilon_{1,2}^q\rightarrow 0$, $R^q$ in Eq. (\ref{eq.RMR}) must  exhibit this chiral  $[SO(2)_L\times SO(2)_R]^q$ symmetry.
In terms of   left-handed $SO(2)_L^{u,d}$ symmetry, the quark mixing matrix  need to be updated into
	\begin{eqnarray}
		V_{CKM0}=R_3(\theta^u_L)R^u{\rm diag}(1, e^{i\lambda_1}, e^{i\lambda_2})(R^d)^TR_3^T(\theta^d_L)
	\label{eq.Vckmh0}
	\end{eqnarray}
Here, the subscribe $0$ stands for hierarchy limit.
Notice that 3-dimensional orthogonal transformation $R^{u}$ and $R^{d}$ need to be determined from  flavor physics models. 
Theoretically, it is possible that $M_0^u$ and $M_0^d$  show  a common organized structure by choosing a special basis in mass hierarchy limit.
By requiring the homology of $M_0^u$ and $M_0^d$, a flat mass matrix (democratic  matrix)   with all elements equalling to 1 arises naturally under the constraint of CKM experiment results \cite{Zhang2021arXiv}
	\begin{eqnarray}
	\frac{1}{m_\Sigma^{u}}M_0^u=\frac{1}{m_\Sigma^{d}}M_0^d=\frac{1}{3}\Array{ccc}{1&1&1\\ 1&1&1\\ 1&1&1}
	\end{eqnarray}
with family mass $m_\Sigma^q=\sum_i m_i^q$.
In fact, the chiral $[SO(2)_L\times SO(2)_R]^q$ symmetry can be  seen from this flat matrix as well. Defining $SO(2)$ rotation along a random direction of ${\bf n}=(n_x, n_y,n_z)$ 
	\begin{eqnarray}
	R_{\bf n}(\theta)=\Array{ccc}{n_x^2(1-c)+c & n_xn_y(1-c)+n_zs & n_xn_z(1-c)-n_ys \\
			n_xn_y(1-c)-n_zs & n_y^2(1-c)+c & n_yn_z(1-c)+n_xs \\
			n_xn_z(1-c)+n_ys & n_yn_z(1-c)-n_xs & n_z^2(1-c)+c }
	\end{eqnarray}
with $c=\cos\theta,s=\sin\theta$. 
It is not hard to find that $M_0^{q}$ is invariant under chiral rotation $R_{111}$ along ${\bf n}=(1,1,1)/\sqrt{3}$
	\begin{eqnarray}
	R_{111}(\theta_L^q)\Array{ccc}{1&1&1\\ 1&1&1\\ 1&1&1}R_{111}^T(\theta_R^q)=\Array{ccc}{1&1&1\\ 1&1&1\\ 1&1&1}
	\end{eqnarray}
$M_0^{q}$ can be diagonalized by $S_0$ 
	\begin{eqnarray}
		\frac{1}{m_\Sigma^{q}}S_0M_0^{q}S_0^T=\Array{ccc}{0 && \\ & 0 & \\ && 1}
	\end{eqnarray}
with 
	\begin{eqnarray}
		{S}_0=\frac{1}{\sqrt{6}}\Array{ccc}{\sqrt{3}& 0&-\sqrt{3}\\ -1& 2 & -1 \\ \sqrt{2} & \sqrt{2} & \sqrt{2}}
		\label{Eq.S0}
	\end{eqnarray}
Substituting $S_0$ into Eq. (\ref{eq.Vckmh0}), 
CKM matrix from  flat mass patterns of up-type and down-type quarks can be expressed as
	\begin{eqnarray}
		V_{CKM0}=R_3(\theta^u_L)S_0{\rm diag}(1, e^{i\lambda_1}, e^{i\lambda_2})S_0^TR_3^T(\theta^d_L)
	\label{eq.CKMh0}
	\end{eqnarray}

Two rotation angles $\theta_L^{u,d}$ provided by $SO(2)_L^{u,d}$ symmetry  play a non-trivial role in CKM mixing. 
If taking $\theta_L^u=\theta_L^d=0$,  a simple relation on mixing parameters $s_{12}$ and $s_{13}$ can be obtained 
	\begin{eqnarray*}
		2s_{12}^2(1-s_{13}^2)=s_{13}^2
	\end{eqnarray*}
Obviously, this relation fails to live up to current CKM experiment results. When these two rotation angles are introduced into Eq. (\ref{eq.CKMh0}), along with phase parameters $\lambda_1$ and $\lambda_2$, $V_{CKM0}$ is parameterized by four quantities, which  just matches the number of CKM mixing parameters precisely. However, before verifying Eq. (\ref{eq.CKMh0}),  perturbations from non-vanishing $h_{23}^q$ must be discussed.

Concerning the second family mass $m_2^q$, we can add 1-order hierarchy correction to $M_0^q$ and define 
	\begin{eqnarray}
	 M_\delta^{q}=M_0^q
		-\frac{3m_\Sigma^{q}}{4}h_{23}^{q}\Array{ccc}{0 & 1 & 0 \\ 1 & 0 &1 \\  0 & 1 & 0}
	\end{eqnarray}
Here, the subscript $\delta$ stands for hierarchy correction.
$M_\delta^{q}$ can be diagonalized  by $S^q_\delta$ 
	\begin{eqnarray}
		S^q_\delta M_\delta^{q}(S^q_\delta)^T=\frac{m_\Sigma^{q}}{3}{\rm diag}\Big(0,h_{23}^q,1\Big)
	\end{eqnarray}
with
	\begin{eqnarray}
		S^q_\delta=S_0+\frac{1}{4\sqrt{3}}h_{23}^q\Array{ccc}{0&0&0\\ \sqrt{2}&\sqrt{2} &\sqrt{2} \\ 1 & -1 & 1}+\mathcal{O}(h^2)
	\end{eqnarray}
Now, the chiral $[SO(2)_L\times SO(2)_R]^q$ symmetry is obviously broken by $h_{23}^q$. However, there is still an approximate $SO(2)_{LR}^q $ symmetry that keeps the eigenvalues of $M_\delta^{q}$ invariant up to $\mathcal{O}(h^1)$.  Defining a rotated $M_\delta^q(\theta)$
	\begin{eqnarray}
		M_\delta^q(\theta)=R^q_\delta(\theta)M^q_\delta [R^q_\delta(\theta)]^T
	\end{eqnarray}
with a  rotation $R_\delta$ along the direction $(1,1-\frac{9}{4}h^q_{23},1)$, 
$M_\delta^q(\theta)$ has the same eigenvalues as $M_\delta^q$
	\begin{eqnarray}
		\Big[S^q_\delta [R^q_\delta(\theta)]^T\Big]M_\delta^q(\theta)\Big[R^q_\delta(\theta)(S^q_\delta)^T\Big]=S^q_\delta M_\delta^q(S^q_\delta)^T=3m_\Sigma^q{\rm diag}(0,h^q_{23},1)+\mathcal{O}(h^2)
	\end{eqnarray}
Here, $R^q_\delta$ includes a perturbative  correction from $h_{23}^q$
	\begin{eqnarray}
		R^q_\delta=R_{111}+
			\frac{h_{23}^q}{4}\Array{ccc}{
				2(1-c_\theta) & -1+c_\theta+\sqrt{3}s& 2(1-c_\theta+\sqrt{3}s_\theta) \\
				-1+c_\theta-\sqrt{3}s_\theta & -4+4c_\theta & -1+c_\theta+\sqrt{3}s_\theta \\
				2(1-c_\theta-\sqrt{3}s_\theta) & -1+c_\theta-\sqrt{3}s_\theta & 2(1-c_\theta)}
	\end{eqnarray}	
Thus, CKM mixing with 1-order hierarchy correction becomes
	\begin{eqnarray}
		V_{CKM}=\Big[S^u_\delta [R^u_{\delta}(\theta^u)]^T\Big]\Big[{\rm diag}(1, e^{i\lambda_1},e^{i\lambda_2})\Big]\Big[R^d_{\delta}(\theta^d)(S^d_\delta)^T\Big]
	\label{eq.ClostToFlatCKM01}
	\end{eqnarray}
Replacing broken chiral $[SO(2)_L\times SO(2)_R]^{q}$ symmetry, $SO(2)_{LR}^{u,d}$ keeps the CKM mixing structure unchanged. 

Assuming that Dirac neutrino has normal-order hierarchal masses, it is not difficult to generalize Eq. (\ref{eq.ClostToFlatCKM01})  to lepton PMNS mixing matrix
\begin{eqnarray}
		U_{PMNS}=\Big[S^e_\delta [R^e_{\delta}(\theta^e)]^T\Big]\Big[{\rm diag}(1, e^{i\lambda_1},e^{i\lambda_2})\Big]\Big[R^\nu_{\delta}(\theta^\nu)(S^\nu_\delta)^T\Big]
	\label{eq.ClostToFlatPMNS01}
	\end{eqnarray}

\begin{figure}
		\centering  
		\includegraphics[scale=0.13]{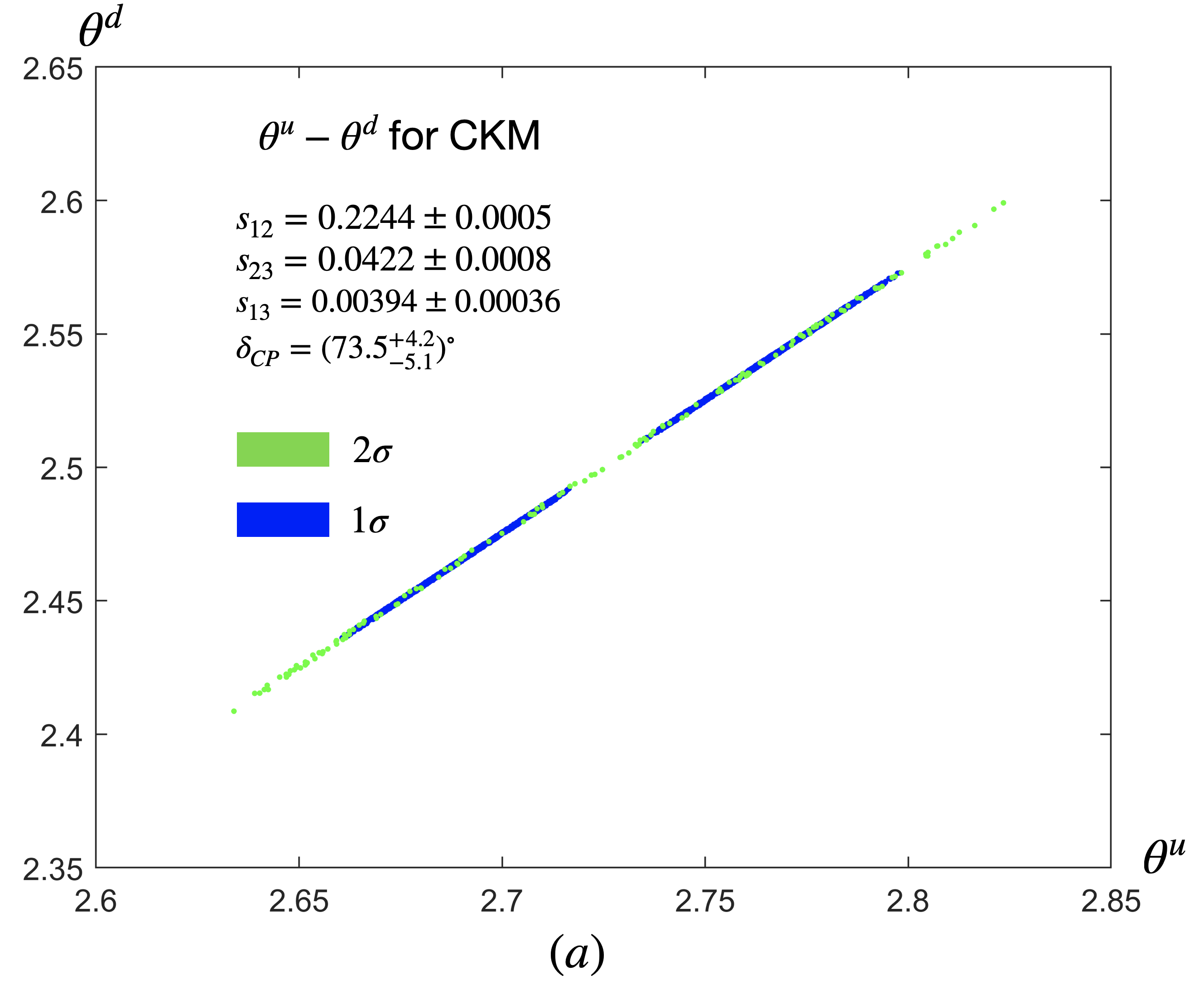}
		\includegraphics[scale=0.13]{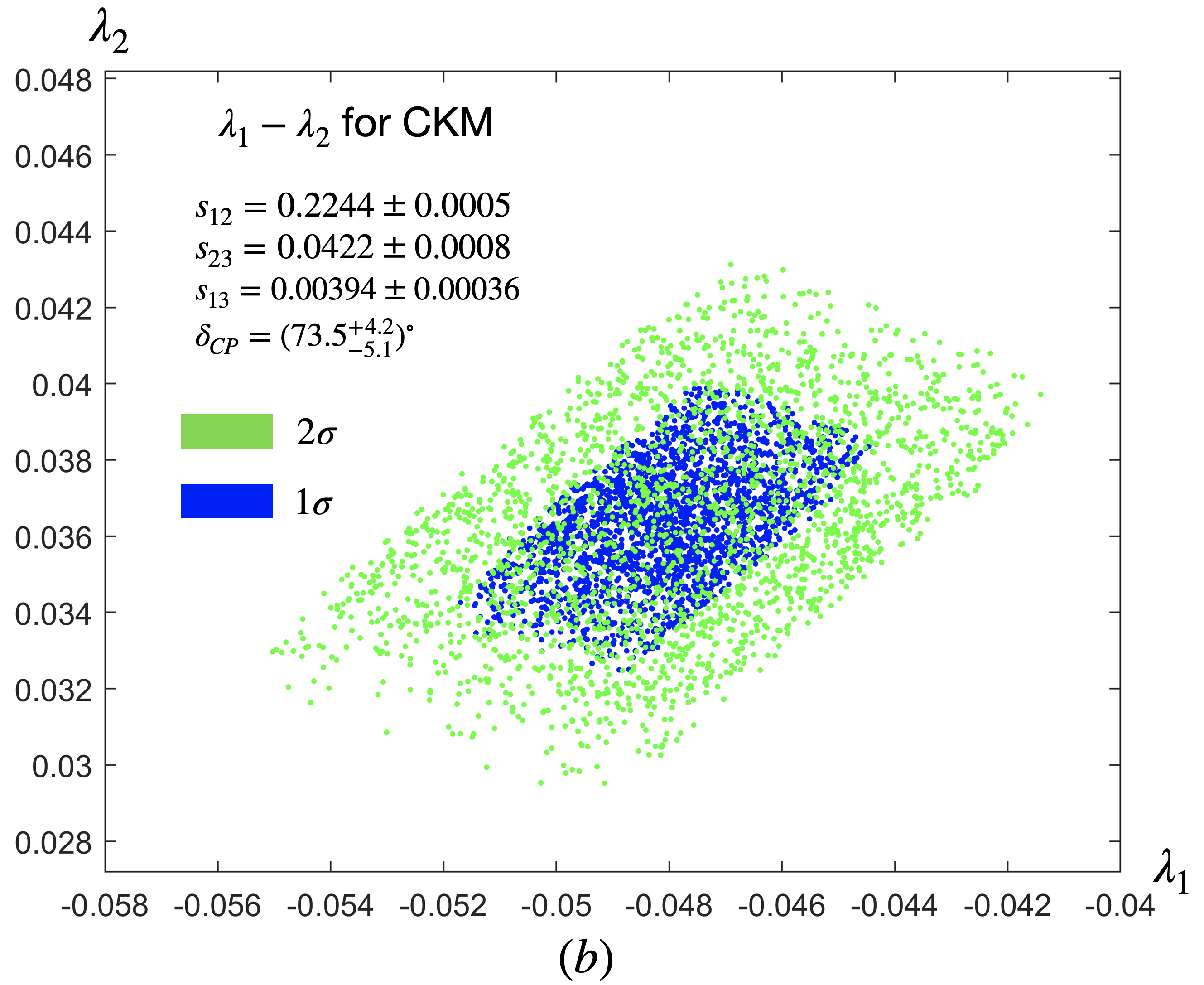}
		\\  
		\includegraphics[scale=0.13]{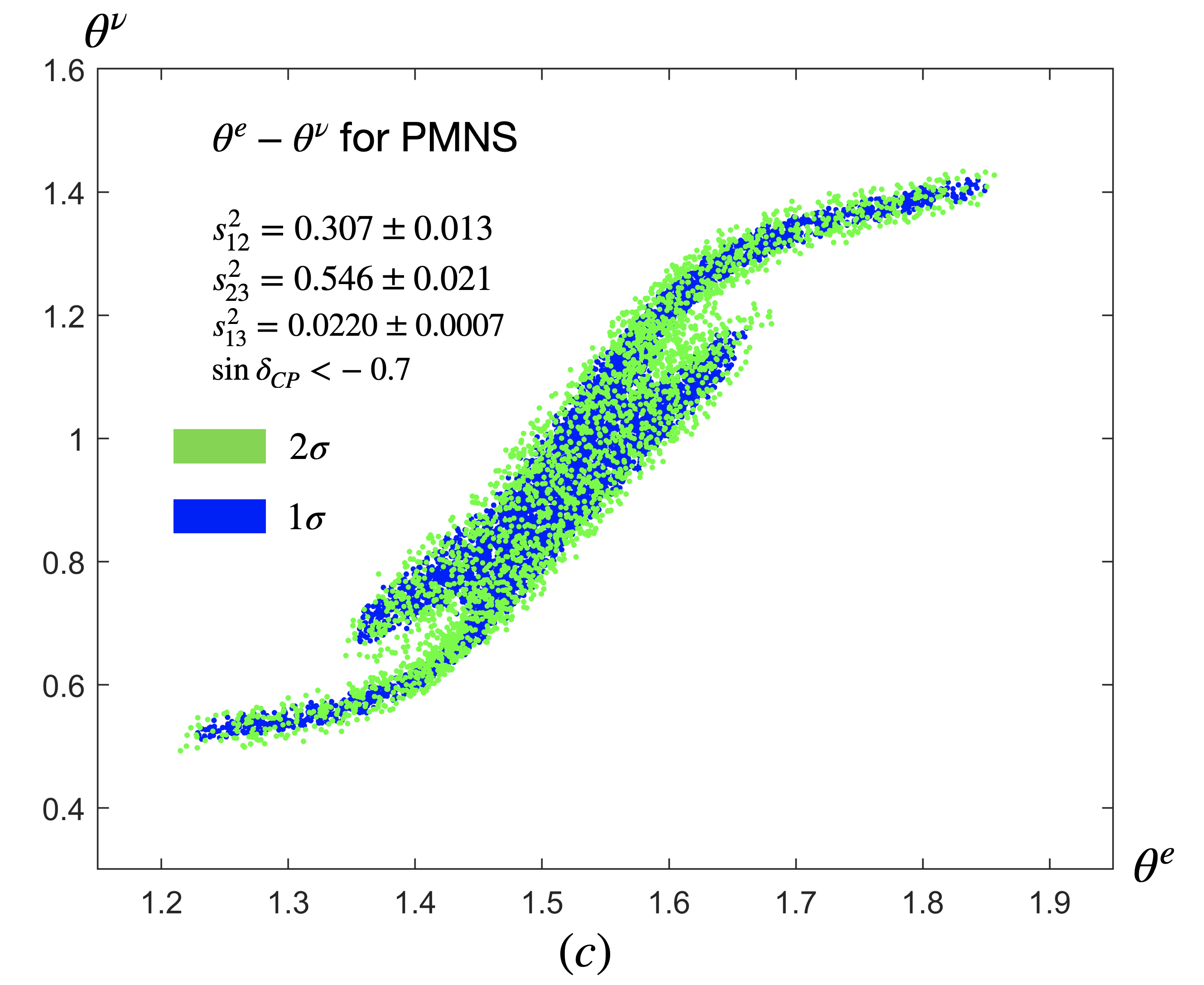}
		\includegraphics[scale=0.13]{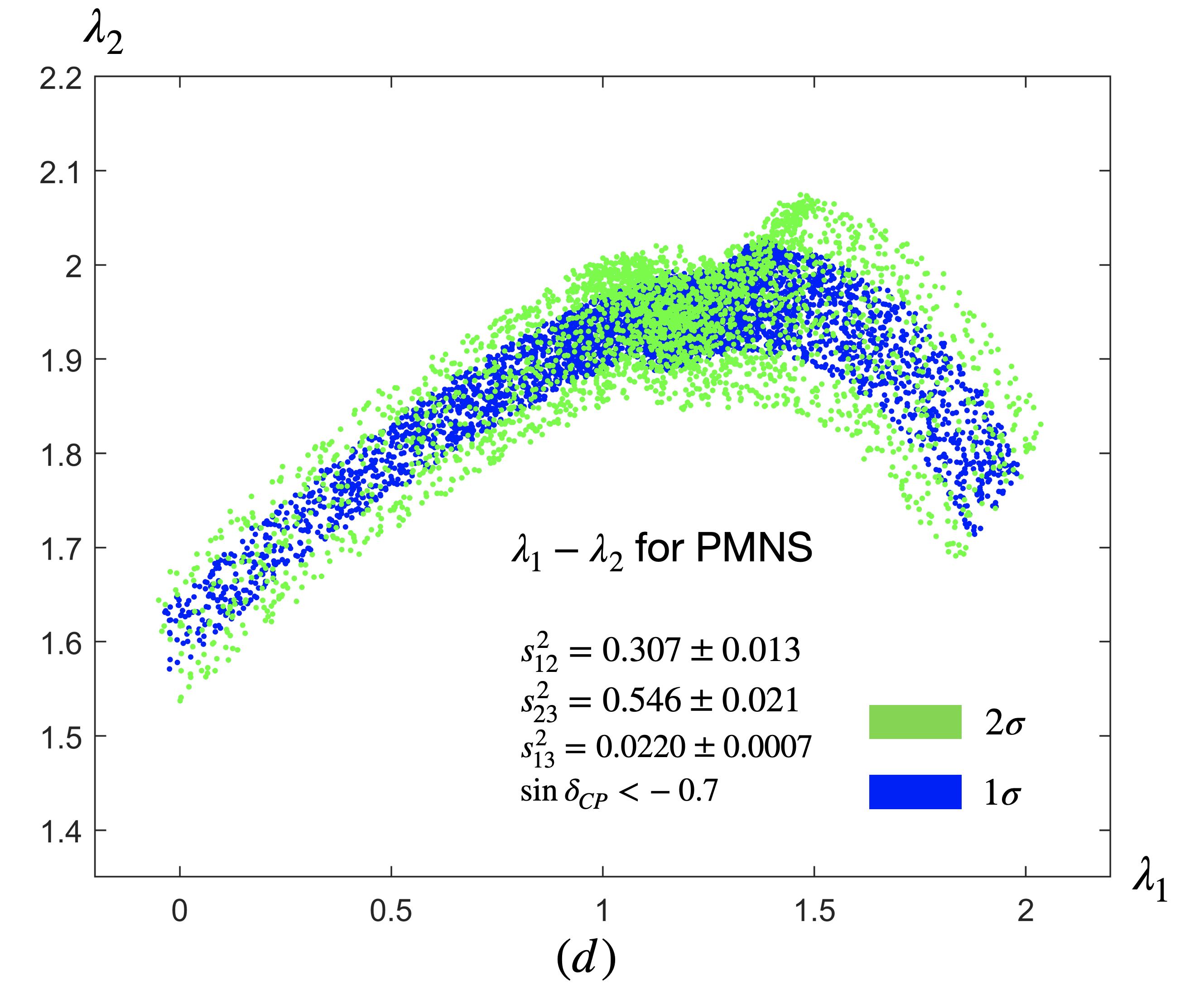}
		\caption{parameter space of $SO(2)_{LR}$ mixing structure for quarks and leptons.}
		\label{fig.fitCKMPMNS}
	\end{figure}

To examine the validity of above  flavor mixing structures in Eq. (\ref{eq.ClostToFlatCKM01}) and Eq. (\ref{eq.ClostToFlatPMNS01}), some calculations have been done. We pick up  four parameters $\theta^u,\theta^d,\lambda_1,\lambda_2$ for quarks  ($\theta^e,\theta^\nu,\lambda_1,\lambda_2$ for leptons) randomly, and put them into Eq. (\ref{eq.ClostToFlatCKM01}) (Eq. (\ref{eq.ClostToFlatPMNS01}) for leptons). Then we extract four CKM (PMNS) mixing parameters and see if it is in the $1\sigma$ ($2\sigma$) C.L. of current CKM/PMNS experiment results \cite{PDG2022}. Plotting all the fitting points gives Fig. \ref{fig.fitCKMPMNS}.
In the conventional view, CKM matrix can be expressed as $V=I+\Delta V$ with a perturbation $\Delta V$. The hierarchal structure of $V$  is attributed to small quark mass ratios. In hierarchy limit, $V$ tend to identity matrix, which equals to vanishing rotation angles $\theta^u=\theta^d=0$. However, from results in Fig. \ref{fig.fitCKMPMNS}(a), the range of $\theta^u$ and $\theta^d$ is not close to the origin, which indicates that the  role of $\theta^{u,d}$ in Eq. (\ref{eq.ClostToFlatCKM01}) can not be neglected. 

Fig. \ref{fig.fitCKMPMNS}(c) and (d) show the fitting points of lepton PMNS mixing, it indicates that we have gotten a same mixing structure for leptons. As we have known, when generalizing traditional sequential seesaw mechanism to lepton, we knock up against an insurmountable difficulty from large PMNS mixing angles $\theta_{12}$ and $\theta_{23}$. A new neutrino mass pattern differing from quark's has to be introduced \cite{Fritzsch2017,BarrPRD2015}.  Now, instead of  a different pattern, 
neutrino  can 
be described in the same close-to-flat mass matrix. 
The common mass structure is a key clue to comprehend quark and lepton flavor in a unified form. 
At a fit point $(\theta^e,\theta^\nu,\lambda_1,\lambda_2)=(1.245,0.5252,1.936,1.760)$, the two large mixing parameters $s_{12}^2$ and $s_{23}^2$ in PMNS matrix can be expressed as
	\begin{eqnarray}
		s_{12}^2&=&0.2773+0.0326-0.0001
		\\
		s_{23}^2&=&0.5500-0.0045+0.0077
	\end{eqnarray}
The first value comes from hierarchy limit result and 
the second and the third are hierarchy corrections from $h_{23}^e$ and $h_{23}^\nu$ (set $h_{23}^\nu=0.172$) , respectively.
This result indicates that two large mixing angles in PMNS parameters can be explained by $SO(2)_{LR}^{e,\nu}$ mixing structure. The lepton mass hierarchies only yield small perturbations.

To summarize, in this paper  an approximate $[SO(2)_L\times SO(2)_R]^f$ (for $f=u,d,e,\nu$) symmetry of mass matrixes of quarks and leptons has been proposed to  address CKM and PMNS mixings in a common fashion. It fits current experiment results successfully and shows that  CKM/PMNS mixing structure comes from an independent mechanism, which is independent from mass hierarchy problem.
Mass hierarchy breaks the chiral $[SO(2)_L\times SO(2)_R]^f$ symmetry to $SO(2)_{LR}^f$ and only provides a slight correction to mixing parameters.
With the help of these results, flavor problems of quark and lepton  can be understood from a unified fashion in the future.

\section*{Acknowledgements}
This work is supported in part by Shaanxi Natural Science Foundation 2022JM-052 and SAFS 22JSY035 of China.

\end{document}